\documentstyle[aps,preprint,prb]{revtex}
\begin{document}
\def\vk{\vec k} 
\def\br{{\bf r}}
\title{\bf A New Approach to the Josephson Effect }
\author{Yong-Jihn Kim}
\address{Department of Physics, University of Puerto Rico$^{\dagger}$ \\
 Mayaguez, PR 00681\\
 Department of Physics,  Bilkent University\\
 06533 Bilkent, Ankara, Turkey}
\maketitle
\begin{abstract}
We introduce a new approach to the Josephson effect
in SIS tunnel junctions.
The Josephson coupling energy is calculated from
the overlap of real space Cooper pair wavefunctions in two 
superconductors through an insulating barrier. 
It is shown that the Josephson tunneling is limited by the 
size of the Cooper pair and its shrinking during the tunneling. 
Therefore, the Josephson coupling energy and the critical 
current become extremely small in high $T_{c}$ superconductors, including $MgB_{2}$. 
This shrinking also causes the observed DC supercurrent in low $T_{c}$ superconductors, 
such as Pb and Sn, to fall off much faster than $1/R_{n}$ for tunneling resistance $R_{n}$ 
above several ohms.
Consequently there is a material-dependent threshold resistance, above which
the supercurrent decreases much faster with increasing resistance.
The impurity-induced shrinking is also shown to limit the critical current.
Furthermore, the (weak) temperature dependence of the Cooper pair size is 
found to contribute to the temperature dependence of the DC supercurrent.  

\end{abstract}
\vskip 3pc
$\dagger$ Present Address

PACS numbers: 74.50.+r, 74.20Fg, 74.25.Sv
\newpage
\section{\bf Introduction} 

The Josephson effect may be the most fascinating and intriguing
property of superconductors.$^{1,2}$
In 1962 Josephson predicted that supercurrents can flow through the 
insulating barrier between two superconductors due to the Cooper pair 
tunneling.$^{1}$ 
The supercurrent depends on the relative phase of two superconductors, where 
the phase is associated with the order parameter of the 
superconducting condensed state. The prediction was confirmed in experiments 
quickly.$^{3,4}$
It is interesting that the experiment by  Nicol et al.$^{5}$ showed already the
possible DC supercurrent, although its reality was not noticed.
This discovery led to the fabrication of the Josephson junction devices and 
the high precision measurement of fundamental constants.$^{6}$  
Now it seems that our understanding of the Josephson effect is on firm
ground.$^{6,7,8}$

Josephson$^{1,2}$ used the tunneling Hamiltonian$^{9}$ to calculate  
tunneling currents between two superconductors and noticed that even at zero 
applied voltages, a DC supercurrent can occur. At finite voltages V, 
he found an AC supercurrent of the same 
amplitude as that of the DC supercurrent, with frequency 2eV/h.
The amplitude of the supercurrent was calculated by Anderson$^{7}$ and 
Ambegaokar and Baratoff.$^{10}$
At $T=0K$ their result of the DC supercurrent, j, is
\begin{equation}
j=j_{1}sin\phi,
\end{equation}
with 
\begin{equation}
j_{1}={\pi\over 2e}{\Delta\over R_{n}}.
\end{equation}
Here $\phi$ is the phase difference of the two superconductors
with the same energy gap $\Delta$.
Note that the amplitude of the supercurrent is proportional to the
energy gap and inversely proportional to the tunneling resistance. 

However, there are still some fundamental experiments which remain puzzling.$^{11,12,13,14,15}$
For instance, the pair-quasi particle interference term was shown to be
negative in experiments, whereas the theory predicts positive sign.$^{11}$
For AC Josephson effect, Likharev even claimed that {\sl there are more questions than answers}.$^{12}$
Furthermore, high $T_{c}$ cuprate Josephson junctions are mainly SNS type 
junctions, since SIS junctions do not show the Josephson effect.$^{13}$
This result is not consistent with the above theory, 
which predicts bigger Josephson current for higher $T_{c}$.$^{1,2,7,8,10}$ 
Recently, this behavior has also been found in $MgB_{2}$ SIS junctions.$^{14,15}$
In particular, the Josephson supercurrent was not observed for the big gap ($\sim 7meV$) of $MgB_{2}$.
We stress that this behavior is closely related to the anomalous dependence of 
the maximum DC Josephson current on the tunneling resistance $R_{n}$, found in 
low $T_{c}$ superconductors, such as Pb and 
Sn:$^{16,17,18,19}$ the Josephson current decreases much faster than $1/R_{n}$ 
above several $\Omega$. 
 
In this paper we present Cooper pair wavefunction approach to 
the Josephson 
effect. It is shown that the Josephson coupling energy is determined by the
overlap of the Cooper pair wavefunctions of two superconductors divided by
a thin insulating layer. (In fact, this idea was suggested by Josephson in his 
original paper,$^{1}$ and confirmed by Ambegaokar and Baratoff.$^{10}$ 
However, it was not pursued thoroughly.)
The critical current is, then, calculated from the coupling energy $E_{J}$ in 
the usual way.$^{2}$ 
Since the Cooper pair tunneling can be understood more easily from the 
Cooper pair wavefunction, this method is advantageous to deal with the 
Josephson effect in high $T_{c}$ superconductors including Mg$B_{2}$ and 
for the insulating barriers with high tunneling resistance $R_{n}$.
We have found that the Cooper pair tunneling is strongly limited by the size 
of the Cooper pair and its reduction during the tunneling. 
Since the Cooper pair size is also reduced by the impurity potential
scattering, ordinary impurity also limits the supercurrent.
As a result, the Josephson current is much smaller than expected in higher 
$T_{c}$ SIS junctions and falls off much faster than $1/R_{n}$ for thicker
and/or higher insulating barriers, in good agreement with the experimental
findings.
The preliminary result was reported before.$^{20}$

\section{\bf Cooper pair wavefunction approach to Josephson Tunneling : T=0K}

\subsection{\bf Josephson Coupling Energy}

The approximate expression of the supercurrent suggested by Josephson$^{1}$ is
\begin{equation}
j\cong{1\over 2}j_{1}\psi^{*}_{l}\psi_{r}+{1\over 2}j_{1}\psi_{r}^{*}\psi_{l},
\end{equation}
where $\psi_{l}$ and $\psi_{r}$ are the effective superconducting 
wavefunctions on the left and right sides, respectively.$^{21,22}$
Actually, it can be exact and it is better to write first the Josephson 
coupling energy $E_{J}$ in terms of the Cooper pair wavefunctions and then 
calculate the supercurrent j from $E_{J}$:$^{2}$ 
\begin{equation}
 j={2e\over \hbar}{\partial E_{J}\over \partial \phi}.
\end{equation}

Figure 1 shows the real space Cooper pair wavefunctions of two superconductors 
with the same energy gap near the insulating barrier. 
The solid lines denote the Cooper pair wavefunctions in
the absence of the barrier, whereas the thick lines denote the change of the
Cooper pair wavefunctions due to the insulating barrier.
The Josephson coupling energy is, therefore, determined by the overlap of the
Cooper pair wavefunctions, connected by the phonon Green's function.$^{23,24}$
Previously, this idea was employed in some cases.$^{25,26}$
For instance, for a pure superconductor without a barrier with the Einstein 
phonons, the coupling energy or the pairing energy is given by
\begin{eqnarray}
E_{int}&=&V\int\int F^{*}(x,y)F(x,y)\delta(x-y)dxdy\nonumber\\
 &=& V\sum_{k}\sum_{k'}u_{k}v_{k}u_{k'}v_{k'},
\end{eqnarray}
where the (effective) Cooper pair wavefunctions $F(x,y)(=\sum_{k}u_{k}v_{k}e^{i{\bf k}\cdot ({\bf r}-{\bf y})}$) and $F^{*}(x,y)$ are coupled by the Einstein
phonon Green's function, i.e., the Dirac delta function, $\delta(x-y)$.
Here V is the phonon-mediated matrix element. 
In the same way, the Josephson coupling energy $E_{J}$ is given by
\begin{eqnarray}
E_{J}&=&V \int\int F_{r}^{*}(x,y)F_{l}(x,y)\delta(x-y)dxdy +V \int\int F_{l}^{*}(x,y)F_{r}(x,y)\delta(x-y)dxdy\nonumber\\
&=& V\int F_{r}^{*}(x)F_{l}(x)dx
+ V\int F_{l}^{*}(x)F_{r}(x)dx,
\end{eqnarray}
with $F_{l}$ and $F_{r}$ being the effective Cooper pair wavefunctions in left 
and right sides.
Observe that this expression has the same form as the approximate one, Eq. (3) 
except the integration.
In fact, Ambegaokar and Baratoff$^{10}$ noted that Josephson coupling 
(free) energy can be calculated by the overlap of Cooper pair wavefunctions, 
which agrees with our result: 
\begin{equation}
E_{J}=\int_{0}^{1}(d\lambda/\lambda )<\lambda H_{T}>,
\end{equation}
where $\lambda$ is an explicit coupling constant and $H_{T}$
denotes the tunneling Hamiltonian.
Figure 1 shows that it is essential to calculate the tail of the Cooper pair 
wavefunctions to determine the Josephson coupling energy.

It is necessary to emphasize that Eq. (6) leads to the familiar result in k-space.
From the tunneling Hamiltonian$^{9}$
\begin{equation}
H_{T}=\sum_{kk's}(T_{kk'}c_{k's}^{r+}c_{ks}^{l}+H.C.)
\end{equation}
one finds the Josephson pair tunneling Hamiltonian, $H_{J}$,$^{27,28,29}$
\begin{equation}
H_{J}=-\sum_{kq}(J_{kq}c^{r+}_{k\uparrow}c^{r+}_{-k\downarrow}c^{l}_{-q\downarrow}c^{l}_{q\uparrow}+H.C.).
\end{equation}
Then, the coupling energy is 
\begin{eqnarray}
E_{J}&=&-J(\sum_{k}u^{r}_{k}v_{k}^{r*}\sum_{q}u_{q}^{l}v_{q}^{l}+
\sum_{k}u^{r}_{k}v_{k}^{r}\sum_{q}u_{q}^{l}v_{q}^{l*})\nonumber\\
&=&-J(\sum_{kq}F_{k}^{r*}F_{q}^{l}+\sum_{kq}F_{k}^{r}F_{q}^{l*})
\end{eqnarray}
with $J_{kq}\equiv J$.
Comparing Eqs. (6) and (10), it is evident that J is proportional to the
phonon-mediated matrix element V, i.e., 
\begin{equation}
J\propto V.
\end{equation}
Accordingly, it is crucial to note that $J_{kq}$ is the pair scattering matrix 
element across the barrier due to both the tunneling and the electron-phonon 
interaction.
If we define 
\begin{eqnarray}
\Psi_{r}&\equiv &\sum_{k}F_{k}^{r}=|\Psi_{r}|e^{i\phi_{r}}\nonumber\\
\Psi_{l}&\equiv &\sum_{q}F_{q}^{l}=|\Psi_{l}|e^{i\phi_{l}}
\end{eqnarray}
Eq. (10) is rewritten
\begin{equation}
E_{J}=-J(\Psi_{l}^{*}\Psi_{r}+\Psi_{l}\Psi_{r}^{*})=-2J|\Psi_{l}||\Psi_{r}|cos(\phi_{l}-\phi_{r}).
\end{equation}

It should be noticed that the result is formally the same as the internal
Josephson coupling energy in two-band superconductors,$^{27,28}$ and believed 
to reproduce the Josephson coupling energy $E_{J}$ in conventional SIS
junctions.$^{27,28,29}$
However, the exact expression of $J_{kq}\equiv J$ is missing, although  
the (2nd order) tunneling-induced pair transfer term ($J^{T}_{kq}$),
\begin{equation}
J^{T}_{kq}={|T_{kq}|^{2}\over E_{k}^{l}+E_{q}^{r}},
\end{equation}
was confused with J.$^{27,29}$
Here $E_{k}^{l}$ and $E_{q}^{r}$ are quasi-particle energies. 
It would be dangerous indeed to suppose that $J_{kq}\equiv J$ (phonon-mediated 
pair-scattering matrix element across the barrier) and $J^{T}_{kq}$ 
(tunneling-induced only pair-scattering matrix element) are the same when they 
are not. Note that $J_{kq}^{T}$ is not proportional to V, unlike J, 
so Eq. (10) or Eq. (13) with $J^{T}_{kq}$ does not reproduce the 
previous result, Eq. (2), which remains puzzling.$^{29}$

\subsection{\bf k-space Approach }

Now we need to compute Eq. (6) or Eq. (10) to find the Josephson coupling 
energy $E_{J}$ and the Josephson current j.
It is clear that both equations should lead to the same result.
First, we consider Eq. (10), which requires the calculation of the 
pair-scattering matrix element $J_{kq}$. For simplicity, we assume two 
superconductors with the same gap, $\Delta$, which will be generalized later. 
From Eq. (9) we obtain 
\begin{equation}
J_{kq}=^{r}<k\uparrow, -k\downarrow|V_{e-ph}|-q\downarrow,q\uparrow>^{l}
\end{equation}
where $V_{e-ph}$ is the phonon-mediated electron-electron interaction.
Tunneling effect is included in the states of the left and right sides. 
It is noteworthy that this situation is much the same as the calculation of 
the matrix element between scattered-state pairs 
in Anderson's theory of dirty superconductors,$^{30,31}$
where impurity effect is included in the scattered states and the 
electron-phonon interaction couples the states.

Using the Tunneling Hamiltonian, Eq. (8), the one-particle wavefunctions are given by
\begin{eqnarray}
\psi^{l}_{q}(r)&=& N_{q}[\phi_{q}^{l}(r)+\sum_{k}{T_{kq}\over \epsilon_{q}-\epsilon_{k}}\phi_{k}^{r}(r)]\nonumber\\
\psi^{r}_{k}(r)&=& N_{k}[\phi_{k}^{r}(r)+\sum_{q}{T_{qk}\over \epsilon_{k}-\epsilon_{q}}\phi_{q}^{l}(r)]
\end{eqnarray}
where $\phi_{q}^{l}$ and $\phi_{k}^{r}$ are the left- and right-hand 
states$^{32}$ with $\epsilon_{q}$ and $\epsilon_{k}$ being the electron 
energies.  $N_{k}$ and $N_{q}$ are the normalization constants.
Prange$^{32}$ proposed the {\sl nonorthogonal almost confined states} for
the left- and right-hand states, whereas Bardeen$^{33}$ introduced the WKB wavefunction,
\begin{eqnarray}
\phi_{m}^{l}&=& Cp_{x}^{-1/2}e^{i(p_{y}y+p_{z}z)} sin(p_{x}x+\gamma), \quad \ \ x<x_{a}\\
\phi_{m}^{l}&=& {1\over 2}C|p_{x}|^{-1/2}e^{i(p_{y}y+p_{z}z)}exp(-\int_{x_{a}}^{x}|p_{x}|dx), \ \ \  x_{a}<x<x_{b}
\end{eqnarray}
where C is a normalization constant and 
$|p_{x}|=(2\mu U-p_{y}^{2}-p_{z}^{2})^{1/2}$ with $U(x)$ being the potential 
energy. $\mu$ is the electron mass. 
The barrier extends from $x_{a}$ to $x_{b}$. 
Notice that the WKB wavefunction is a sinusoidal
stationary wave (in x-direction).
Thus, we have degenerate pairs $m=(p_{x},p_{y},p_{z})$ and 
${\bar m}=(p_{x},-p_{y},-p_{z})$.

For the Einstein phonon model,$^{23,24,31}$ the pair-scattering matrix element is
\begin{eqnarray}
J_{kq}&=&V\int dr \psi_{k}^{r*}(r)\psi_{-k}^{r*}(r)\psi_{-q}^{l}(r)\psi_{q}^{l}(r)\nonumber\\
&=& V\int dr |\psi_{k}^{r}(r)|^{2}|\psi_{q}^{l}(r)|^{2},
\end{eqnarray}
which denotes the (density) correlation function between the eigenstates 
$\psi_{q}^{l}(r)$ and $\psi_{k}^{r}(r)$.   
Since the Bardeen's wavefunction is not much different from a 
plane wave, we may employ plane waves for $\phi_{q}^{l}$ and $\phi_{k}^{r}$ 
to obtain the matrix element,
\begin{equation}
J_{kq}\cong V{1\over \Omega_{r}}\sum_{k}{|T_{kq}|^{2}\over (\epsilon_{q}-\epsilon_{k})^{2}}+
V{1\over \Omega_{l}}\sum_{q}{|T_{qk}|^{2}\over (\epsilon_{k}-\epsilon_{q})^{2}},
\end{equation}
where $\Omega_{l}$ and $\Omega_{r}$ are the volumes of the left and right sides.
Unfortunately, the sums are divergent, which are common in the perturbation 
theory for continuous spectra. The remedy is to use the scattering theory 
with a cutoff ($\xi_{0}$), consistent with the bound state of Cooper 
pairs.$^{34}$
Since each sum denotes the relative probability contained in the virtual 
scattered wavelets (due to the barrier) within the BCS coherence length $\xi_{0}=\hbar v_{F}/\pi\Delta$ 
(compared to the plane wave part),$^{34}$ we find
\begin{eqnarray}
{1\over \Omega_{r}}\sum_{k}{|T_{kq}|^{2}\over (\epsilon_{q}-\epsilon_{k})^{2}}&\approx & <|T_{kq}|^{2}>{N_{F}\pi^{2}\over 2\hbar v_{F}}\xi_{0} = <|T_{kq}|^{2}>{N_{F}\pi\over 2\Delta}\nonumber\\
{1\over \Omega_{l}}\sum_{q}{|T_{qk}|^{2}\over (\epsilon_{k}-\epsilon_{q})^{2}}&\approx & <|T_{qk}|^{2}>{N_{F}\pi^{2}\over 2\hbar v_{F}}\xi_{0} = <|T_{qk}|^{2}>{N_{F}\pi\over 2\Delta}
\end{eqnarray}
and
\begin{equation}
<J_{kq}>\equiv J\approx V<|T|^{2}>{N_{F}\pi\over \Delta}.
\end{equation}
Here $N_{F}$ is the density of states at the Fermi level and the angular brackets indicate the average value over the states.

Consequently, substituting Eq. (22) to Eq. (13) the Josephson coupling energy is written
\begin{equation}
E_{J}\approx - {2\pi\over \lambda} <|T|^{2}>N_{F}^{2}\Delta\ cos(\phi_{l}-\phi_{r}),
\end{equation}
where $\lambda=N_{F}V$.
Accordingly, the Josephson supercurrent is
\begin{equation}
j={1\over \lambda}{\Delta\over eR_{n}}\ sin(\phi_{l}-\phi_{r}),
\end{equation}
with
\begin{equation}
j_{1}={1\over \lambda}{\Delta\over eR_{n}}.
\end{equation}
Although this equation for the  maximum DC supercurrent is similar to the
previous one, Eq. (2), physically it is significantly different. The factor
$1/\lambda$ shows that the Josephson coupling energy and the supercurrent
depend on the superconductor. (Notice that this factor also appears in the 
initial $T_{c}$ decreases due to magnetic impurities$^{34}$ and weak 
localization.$^{31}$)
As a result, if we include the reduction of the Cooper pair size during the 
tunneling, which is important especially for superconductors with high $T_{c}$, 
the resulting supercurrent is not similar at all. 

When the superconductors have different energy gaps, $\Delta_{1}$ and 
$\Delta_{2}$, the pair-scattering matrix elements are 
\begin{equation}
<J_{kq}>\equiv J\approx V_{1}<|T|^{2}>{N_{F,1}\pi\over 2\Delta_{2}}
+ V_{2}<|T|^{2}>{N_{F,2}\pi\over 2\Delta_{1}}
\end{equation}
where $N_{F,1}$ and $N_{F,2}$ are the density of states of left and
right sides, respectively.
Subsequently, one finds
\begin{equation}
E_{J}= - \pi <|T|^{2}>N_{1,F}N_{F,2}\Delta_{1}\Delta_{2}({1\over \lambda_{1}\Delta_{1}}+{1\over \lambda_{2}\Delta_{2}}) cos(\phi_{l}-\phi_{r}),
\end{equation}
and
\begin{equation}
j={2\pi\over \hbar}e<|T|^{2}>N_{F,1}N_{F,2}\Delta_{1}\Delta_{2} ({1\over \lambda_{1}\Delta_{1}}+{1\over \lambda_{2}\Delta_{2}})  sin(\phi_{l}-\phi_{r}),
\end{equation}
with
\begin{equation}
j_{1}={1\over 2e}{1\over R_{n}}\Delta_{1}\Delta_{2} ({1\over \lambda_{1}\Delta_{1}}+{1\over \lambda_{2}\Delta_{2}}) .
\end{equation}
Here $\lambda_{1}$ and $\lambda_{2}$ are the BCS coupling constants for two superconductors.

\subsection{\bf Cooper Pair Wavefunction Approach}

We consider now Eq. (6), the Josephson coupling energy in terms of the Cooper
pair wavefunctions. This method is more appropriate for high $T_{c}$ 
superconductors, including $MgB_{2}$, which have tightly-bound Cooper pairs 
with small size. It is also desirable for the insulating barrier with high 
tunneling resistance $R_{n}$, since the Cooper pair size shrinks significantly 
during the tunneling in that case. A similar reduction of the Cooper pair size 
caused by the ordinary impurity scattering requires this approach, too.

The Cooper pair wavefunctions are expressed in terms of the one-particle 
wavefunctions, Eq. (16):
\begin{eqnarray}
F_{l}(x)&=& \sum_{q}u_{q}v_{q}\psi_{q}^{l}(x)\psi_{-q}^{l}(x)\\
F_{r}(x)&=& \sum_{k}u_{k}v_{k}\psi_{k}^{r}(x)\psi_{-k}^{r}(x).
\end{eqnarray}
Substituting Eqs. (30) and (31) into Eq. (6), it is straightforward to show that
\begin{equation}
E_{J}=V( {1\over \Omega_{r}}\sum_{k}{|T_{kq}|^{2}\over (\epsilon_{q}-\epsilon_{k})^{2}}+ {1\over \Omega_{l}}\sum_{q}{|T_{qk}|^{2}\over (\epsilon_{k}-\epsilon_{q})^{2}})(\sum_{kq}u_{k}^{r}v_{k}^{r*}u_{q}^{l}v_{q}^{l}+\sum_{kq}u_{q}^{l}v_{q}^{l*}u_{k}^{r}v_{k}^{r})
\end{equation}
which is indeed the same as that obtained by the k-space approach, i.e., 
Eq. (10) with Eq. (20).
Whereas this real space approach allows us to take into account the change of 
the Cooper pair size during the tunneling through the Cooper pair wavefunctions.

If we insert the Bardeen's WKB wavefunction into Eq. (30), 
the left-side Cooper pair wavefunction shows the exponential decay in the 
barrier region:
\begin{equation}
F_{l}(x)={C^{2}\over 4}(\sum_{p}u_{p}v_{p})|p^{0}_{x}|^{-1}e^{-2\kappa(x-x_{a})} \quad x_{a}<x<x_{b}
\end{equation}
where $|p^{0}_{x}|=\hbar\kappa=(2\mu U_{0}-p_{y}^{2}-p_{z}^{2})^{1/2}$ with
$U_{0}$ being the potential energy.
Beyond $x_{b}$, the tail of the Cooper pair wavefunction should be, after utilizing the one-particle wavefunction, Eq. (16), 
\begin{equation}
F_{l}(x)=\sum_{q}u_{q}v_{q}[\sum_{k}{|T_{kq}|^{2}\over (\epsilon_{q}-\epsilon_{k})^{2}}\phi_{k}^{r}(x)\phi_{-k}^{r}(x)] \quad x>x_{b}.
\end{equation}
We have used the fact that only ($\phi_{k}^{r}$, $\phi_{-k}^{r}$) pair 
contributes to the coupling energy, Eq. (32). (It is obvious that 
$|T_{kq}|^{2}\propto e^{-2\kappa d}$, with the barrier thickness 
$d=x_{b}-x_{a}$.$^{35,36}$)
However, since $\phi_{k}^{r}$ and $\phi_{-k}^{r}$ are basically plane-wave-like,
Eq. (34) shows not the presumed exponential tail but a constant 
amplitude. Of course, the divergence in the sum implies this behavior needs to 
be corrected according to the bounded Cooper pair wavefunction.
Therefore, we assume 
\begin{equation}
F_{l}(x)\approx <|T_{kq}|^{2}>\sum_{q}u_{q}v_{q}{N_{F}\pi^{2}\over 2\hbar v_{F}}e^{-(x-x_{b})/\xi_{0}} \quad x>x_{b}.
\end{equation}
(In fact, $x$ denotes the center of mass coordinate of the Cooper pair rather 
than the relative coordinate. Since the phonon-mediated attraction between the
electrons in the left-hand side can propagate only up to the range of the 
Cooper pair size $\xi_{0}$ in the right-hand side, this assumption seems to be
reasonable.)
Inserting Eq. (35) into the first integral in Eq. (6) we obtain (in the right side)
\begin{eqnarray}
V\int^{r} F_{r}^{*}(x)F_{l}(x)dx &=& V<|T|^{2}>\sum_{k}u_{k}^{r}v_{k}^{r*}\sum_{q}u^{l}_{q}v^{l}_{q}{N_{F}\pi^{2}\over 2\hbar v_{F}}\int_{x_{b}}^{\infty}e^{-(x-x_{b})/\xi_{0}}dx\nonumber\\
&=& V<|T|^{2}>\sum_{k}u_{k}^{r}v_{k}^{r*}\sum_{q}u^{l}_{q}v^{l}_{q}{N_{F}\pi\over 2\Delta}
\end{eqnarray}
where $\int^{r}$ denotes the integration over the right side and $x_{a}=0$. 
$F_{r}(x)$ will have the same exponential tail in the left side. 
As expected, the Cooper pair wavefunction method leads to the same result 
for the Josephson coupling energy and the supercurrent:
\begin{eqnarray}
E_{J}&=& -{2\pi\over \lambda} <|T|^{2}>N_{F}^{2}\Delta \ cos(\phi_{l}-\phi_{r})\\
j &=&{1\over \lambda}{\Delta\over eR_{n}}\ sin(\phi_{l}-\phi_{r}),
\end{eqnarray}
where 
\begin{equation}
j_{1}={1\over \lambda}{\Delta\over eR_{n}}.
\end{equation}

\subsection{\bf Shrinking of the Cooper Pair Size Due to the Insulating Barrier}

In the previous section we assumed that the tail of $F_{l}(x)$ in the right 
side, Eq. (34), is controlled by the same Cooper pair size, $\xi_{0}$, of 
the left side. In other words, we disregarded the effect of the insulating 
barrier potential on the Cooper pair size. However, it is well-known that the 
Cooper pair size is reduced due to the impurity potentials.$^{37,38}$ 
Therefore, it is obvious that this assumption is not appropriate especially 
for high $T_{c}$ superconductors, including $MgB_{2}$, and for the insulating 
barriers with high tunneling resistance, $R_{n}$, 
since the Cooper pair will have so much difficulty in tunneling in these cases. 

We estimate the reduction of the Cooper pair size due to the barrier potential.
Equation (33) implies that the decay length of the Cooper pair wavefunction in 
the insulating barrier region is $\sim 1/2\kappa$. 
Whereas, after tunneling the Cooper pair (of size $\xi_{0}$) will have the 
memory of the phonon-mediated attraction up to $\xi_{0}-d$.
So we suppose that the Cooper pair have size $1/2\kappa$ with the weight
$d/\xi_{0}$ and $\xi_{0}$ with the weight $(\xi_{0}-d)/\xi_{0}$.
Accordingly, the effective Cooper pair size $\xi_{eff}$ may be written as
\begin{equation}
{1\over \xi_{eff}}= {1\over \xi_{0}}{\xi_{0}-d\over \xi_{0}}+2\kappa {d\over\xi_{0}}.
\end{equation}
Since $d<<\xi_{0}$ in most cases, we find the effective Cooper pair size in 
the right side,
\begin{equation}
\xi_{eff} \cong {\xi_{0}\over 1+2\kappa d}.
\end{equation}
Then, the effective Cooper pair wavefunction of $F_{\ell}(x)$ in the right side is
\begin{equation}
F_{l}(x)\propto e^{-{x\over \xi_{eff}}}=e^{-{1+2\kappa d\over \xi_{0}}x} \quad\quad x\geq x_{a}=0.
\end{equation}
Now, it is important to notice that this reduction of the Cooper pair size is 
fully meaningful only when the exponential factor in Eq. (42) is $\sim e^{-1}$ just after tunneling, i.e., 
\begin{equation}
F_{l}(d)\propto e^{-{d\over \xi_{eff}}}=e^{-{1+2\kappa d\over \xi_{0}}d}\sim e^{-1}.
\end{equation}
Thus, the Cooper pair size shrinks significantly when 
$d\sim \sqrt{\xi_{0}/2\kappa}$, i.e., for high 
$T_{c}$ superconductors, such as $MgB_{2}$ and for the insulating barriers 
with high $R_{n}$.
To put it another way, there is a (sample-dependent) threshold tunneling 
resistance, $R_{th}$, above which the supercurrent decreases much faster, 
although it is hard to determine the exact value of $R_{th}$.

The overlap of the Cooper pair wavefunctions is given by
\begin{eqnarray}
V\int^{r} F_{r}^{*}(x)F_{l}(x)dx &\cong & V<|T|^{2}>\sum_{k}u_{k}^{r}v_{k}^{r*}\sum_{q}u^{l}_{q}v^{l}_{q}{N_{F}\pi^{2}\over 2\hbar v_{F}}\int_{x_{b}}^{\infty} e^{-{1+2\kappa d\over \xi_{0}}x} dx\nonumber\\
&\cong & V<|T|^{2}>\sum_{k}u_{k}^{r}v_{k}^{r*}\sum_{q}u^{l}_{q}v^{l}_{q}{N_{F}\pi\over 2\Delta}{1\over 1+2\kappa d}e^{-{1+2\kappa d\over \xi_{0}}d}.
\end{eqnarray}
Consequently, for $d\sim \sqrt{\xi_{0}/2\kappa}$ the Josephson coupling energy and the supercurrent are written as
\begin{equation}
E_{J}= - {2\pi\over \lambda} <|T|^{2}>N_{F}^{2}\Delta {1\over 1+2\kappa d} e^{-{1+2\kappa d\over \xi_{0}}d}\ cos(\phi_{l}-\phi_{r}).
\end{equation}
and
\begin{equation}
j= {1\over \lambda}{\Delta\over eR_{n}}{1\over 1+2\kappa d} e^{-{1+2\kappa d\over \xi_{0}}d}\ sin(\phi_{l}-\phi_{r}),
\end{equation}
with
\begin{equation}
j_{1}= {1\over \lambda}{\Delta\over eR_{n}}{1\over 1+2\kappa d} e^{-{1+2\kappa d\over \xi_{0}}d}.
\end{equation}
Since the DC Josephson current is very small in this case, 
fabrication of $MgB_{2}$ SIS junctions requires
very accurate nanometer scale manipulation of the insulating barriers. 
For instance, if $\xi_{0}\sim 100\AA$, we need an insulating barrier of
thickness d smaller than $\sim 10\AA$ (i.e., $R_{th}\sim 0.01-0.1\Omega)$ to see 
the Josephson supercurrent,
which explains why the previous experiments couldn't see the DC supercurrent
corresponding to the big gap of $MgB_{2}$.$^{14,15}$

\subsection{\bf  $R_{n}$ Dependence of the DC Supercurrent }

We have found the maximum DC supercurrent:
\begin{eqnarray}
j_{1}&=& {1\over \lambda}{\Delta\over eR_{n}}\hspace{1in} \quad\quad\ \ \rm{for}\ \rm{ low}\ R_{n}\\
j_{1}&=&{1\over \lambda}{\Delta\over eR_{n}}{1\over 1+2\kappa d} e^{-{1+2\kappa d\over \xi_{0}}d} \quad\ \  \rm{for}\ \rm{ high}\ R_{n}\ (d \sim \sqrt{\xi_{0}/2\kappa}) 
\end{eqnarray}
where 
\begin{equation}
R_{n}^{-1}= {4\pi e^{2}\over \hbar}N_{F}^{2}<|T|^{2}>.
\end{equation}
For low $R_{n}$ the DC supercurrent is inversely proportional to $R_{n}$, 
which may be called {\sl linear region},
whereas for high $R_{n}$, corresponding to $d \sim \sqrt{\xi_{0}/2\kappa}$, 
the supercurrent decreases more quickly with increasing $R_{n}$, which may be 
called {\sl steep region}, although the boundary is not that sharp. 
This behavior agrees with the experiments.$^{15-19}$  
(Presumably in the {\sl steep region} the supercurrent may drop to zero 
exponentially. This is the reason why we kept the exponential factor in the 
previous section.) One needs an interpolation formula that smoothly connects 
both limits, for low $R_{n}$ and for high $R_{n}$. (Nevertheless, high $R_{n}$ 
formula Eq. (49) is 
valid in a rather narrow range where $d\sim \sqrt{\xi_{0}/2\kappa}$ and the 
exponential factor $\sim e^{-1}$.) For that purpose, we may multiply 
$2\kappa d$ in Eq. (49) by the approximate function for the Heaviside unit step function,$^{39}$ i.e., 
\begin{equation}
S_{n}(\kappa d)={1\over 2}[1+tanh\ n(2\kappa d-2\kappa d_{c})]
\end{equation}
where $2\kappa d_{c}$ may be chosen to give the exponential factor equal to 
$e^{-1}$, while n is selected to have the resistance spread of factor of 10, 
according to the experimental data.
(In this case $2\kappa d_{c}$ gives much higher resistance than the threshold resistance $R_{th}$.)

In the presence of ordinary impurities, the Cooper pair size also decreases
from $\xi_{0}$ to $\tilde{\xi_{0}}$, defined by$^{21,31,38}$
\begin{equation}
{1\over \tilde{\xi_{0}}}= {1\over \xi_{0}}+{1\over \ell},
\end{equation}
where $\ell$ is the mean free path.
(It seems that Eq. (52) works better than the other expression, i.e., $\sqrt{\ell\xi_{0}}$.$^{37}$)
The effective Cooper pair size is then replaced by
\begin{equation}
\xi_{eff} \cong {\ell\xi_{0}\over (1+2\kappa d)(\ell+\xi_{0})}.
\end{equation}
It is remarkable that ordinary impurities also decrease the supercurrent
by reducing the Cooper pair size.

Figure 2 shows the comparison of our theoretical calculations with the 
experimental data for Pb-Pb$\rm{O_{x}}$-Pb (PPP) at 4.2K by Schwidtal and 
Finnegan,$^{16}$ and Sn-SnO-Pb (SSP) at 1.4K by Tinkham's group 
(Danchi et. al.).$^{19}$ The coherence lengths $\xi_{0}$ for Pb and Sn are 
$\sim 820\AA$ and $\sim 1800\AA$, respectively.$^{40}$ 
Note that above $R_{n}\sim 40\Omega$ 
the supercurrent of Sn-SnO-Pb junction becomes larger than that of
Pb-Pb$\rm{O_{x}}$-Pb junction, due to the bigger Cooper pair
size of Sn.
For {\sl linear region} $j_{1}=1.5/R_{n}(mA)$ for PPP and
$j_{1}=1.2/R_{n}(mA)$ for SSP.
McMillan and Rowell$^{41}$ observed that for a typical junction with a 
resistance of 30$\Omega$, the exponential factor is roughly $exp(-20)$. 
So the tunneling resistance is assumed to be 
$R_{n}=30\times exp(2\kappa d-20)(\Omega)$. 
The threshold resistance for PPP is $R_{th} \sim 30\Omega$, (i.e., $d_{th}\sim 20\AA$), whereas
for SSP, there are two threshold resistances for Sn and PB, leading to 
(average) value, $R_{th} \sim 50\Omega$.  For PPP junction we used 
$Sn(\kappa d)={1/2}[1+tanh\ 1.9(\kappa d-21.25)]$, whereas for   
SSP junction  we used $Sn(\kappa d)={1/2}[1+tanh\ 1.0(\kappa d-21.25)]$ 
for Pb and $Sn(\kappa d)={1/2}[1+tanh\ 1.0(\kappa d-22.75)]$ for Sn, 
respectively. Unfortunately, the mean free path $\ell$ is not available for 
those experimental data. However, since the thickness of the film is about 
$2,000\AA$,$^{16}$ we suppose $\ell\sim 1000\AA$.
If we assume $\ell=1000\AA$ for Pb-Pb$\rm{O_{x}}$-Pb junction, we obtain 
the impurity-limited coherence length 
$\tilde \xi_{0}=\ell\xi_{0}/(\ell+\xi_{0})=450.6\AA$, leading to
$d_{c}\sim 21.23\AA$, in agreement with the above $R_{n}(\kappa d)$.
For SSP junction, the coherence length of Sn, $\sim 1800\AA$ requires 
$\ell\sim 800\AA$ to get $d_{c}\cong 23.5\AA$, which is also consistent with
the $R_{n}$ used. 
Considering the uncertainty in $\ell$, the perfect fit with the experimental 
data is rather fortuitous. Nevertheless, our approach explains clearly why 
the Cooper pair of Sn can tunnel through the barrier more easily than that of Pb.

\section{\bf Josephson Effect at Finite Temperature}

Now we consider the Josephson coupling energy and the supercurrent at finite
temperatures for low tunneling resistance $R_{n}$. The Bogoliubov-Valatin 
transformations are defined by$^{42}$
\begin{eqnarray}
c_{k\uparrow}^{r}&=&u_{k}^{r}e_{k\uparrow}+v_{k}^{r}e^{+}_{-k\downarrow}\nonumber\\
c_{k\downarrow}^{r}&=&u_{k}^{r}e_{k\downarrow}-v_{k}^{r}e^{+}_{k\uparrow}\\
c_{q\uparrow}^{l}&=&u_{q}^{l}f_{q\uparrow}+v_{q}^{l}f^{+}_{-q\downarrow}\nonumber\\
c_{q\downarrow}^{l}&=&u_{q}^{l}f_{q\downarrow}-v_{q}^{l}f^{+}_{q\uparrow}.
\end{eqnarray}
Consequently, the coupling energy is written
\begin{eqnarray}
E_{J}&=&-J\sum_{kq}[u^{r}_{k}v_{k}^{r*}(1-2f_{k})u_{q}^{l}v_{q}^{l}(1-2f_{q})+ u^{r}_{k}v_{k}^{r}(1-2f_{k})u_{q}^{l}v_{q*}^{l}(1-2f_{q})]\nonumber\\
&=&-2J\sum_{kq}{\Delta_{1}(T)\Delta_{2}(T)cos(\phi_{l}-\phi_{r})\over 2E_{k}^{r}\times 2E_{q}^{l}}(1-2f_{k})(1-2f_{q}).
\end{eqnarray}
At $T=0K$ this expression leads to Eq. (13).
After summations over k and q, we obtain
\begin{equation}
E_{J}=-2J{\Delta_{1}(T)\Delta_{2}(T)\over V_{1}V_{2}}cos(\phi_{l}-\phi_{r}).
\end{equation}
Notice that the pair-scattering matrix element, J, has rather weak temperature 
dependence through the effective Cooper pair sizes $\xi_{eff,1}(T)$ and $\xi_{eff,2}(T)$:
\begin{equation}
J= V_{1}<|T|^{2}>{N_{F,1}\pi^{2}\over 2\hbar v_{F}}\xi_{eff, 2}(T)
+ V_{2}<|T|^{2}>{N_{F,2}\pi^{2}\over 2\hbar v_{F}}\xi_{eff, 1}(T).
\end{equation}
It is well-known that the pair-correlation amplitude, or the Cooper pair 
wavefunction falls exponentially:$^{21,26}$
\begin{eqnarray}
F(r)&\propto& exp(-{r\over \pi\xi_{0}})\quad\quad \quad T=0K\\
F(r)&\propto& exp(-{2r\over 3.5\xi_{0}})\quad \quad\ T\rightarrow T_{c} 
\end{eqnarray}
leading to the decrease of the effective Cooper pair size $\xi_{eff}(T)$ 
from $\pi\xi_{0}/2$ at $T=0K$ to $3.5\xi_{0}/4$ near $T_{c}$.
(In the previous sections we used $\xi_{0}$ instead of $\pi\xi_{0}/2$ at 
$T=0K$.)
The limiting behavior of the effective Cooper pair size can be shown to be$^{43}$ 
\begin{eqnarray}
\xi_{eff}(T)&\cong& {\pi\xi_{0}\over 2}(1-{\pi^{2}T^{2}\over 2\Delta^{2}}) \quad \quad\quad\quad\ \rm{near }\ \ T=0K\\
\xi_{eff}(T) &\cong&{1\over 2} {\hbar v_{F}\over \pi T }(1-{\Delta^{2}\over 2\pi^{2}T^{2}})\quad\quad\quad \rm{near}\ \ T_{c}, 
\end{eqnarray}
where $\xi_{eff}(T)\cong {\pi\Delta_{0}\xi_{0}\over 2\sqrt{\pi^{2}T^{2}+\Delta^{2}}}$ at any temperature.
Unlike $\Delta(T)$, near $T=0K$, $\xi_{eff}(T)$ decreases with increasing
temperature in a parabolic manner.

Accordingly, for a symmetric junction the Josephson coupling energy and the supercurrent are written as
\begin{eqnarray}
E_{J}&=&-2|T|^{2}{\pi^{2}\over \hbar v_{F}}{\xi_{eff}(T)\over \lambda}N_{F}^{2}\Delta^{2}(T)cos\phi\\
j&=&{1\over \lambda}{\pi\xi_{eff}(T)\over \hbar v_{F}}{\Delta^{2}(T)\over eR_{n}}sin\phi,
\end{eqnarray}
with 
\begin{equation}
j_{1}={1\over \lambda}{\pi\xi_{eff}(T)\over \hbar v_{F}}{\Delta^{2}(T)\over eR_{n}}.
\end{equation}
The reduced dc Josephson current is, then, given by
\begin{eqnarray}
{j_{1}(T)\over j_{1}(0K)}&=&{\xi_{eff}(T)\over {{\pi\over 2}\xi_{0}}}{\Delta^{2}(T)\over \Delta_{0}^{2}}\\
&=& {\Delta^{2}(T)\over \Delta_{0}\sqrt{\pi^{2}T^{2}+\Delta^{2}(T)}}.
\end{eqnarray}
	
For an asymmetric junction, one finds the coupling energy and the supercurrent: 
\begin{equation}
E_{J}=- \pi <|T|^{2}>N_{F,1}N_{F,2}\Delta_{1}(T)\Delta_{2}(T) ({\pi\xi_{eff,1}(T)\over \lambda_{1}\hbar v_{F,1}}+{\pi\xi_{eff,2}(T)\over \lambda_{2}\hbar v_{F,2}}) cos(\phi_{l}-\phi_{r}),
\end{equation}
and
\begin{equation}
j={2\pi\over \hbar}e<|T|^{2}>N_{F,1}N_{F,2}\Delta_{1}(T)\Delta_{2}(T)({\pi\xi_{eff,1}(T)\over \lambda_{1}\hbar v_{F,1}}+{\pi\xi_{eff,2}(T)\over \lambda_{2}\hbar v_{F,2}}) sin(\phi_{l}-\phi_{r}),
\end{equation}
with
\begin{equation}
j_{1}={1\over 2e}{1\over R_{n}}\Delta_{1}(T)\Delta_{2}(T) ({\pi\xi_{eff,1}(T)\over \lambda_{1}\hbar v_{F,1}}+{\pi\xi_{eff,2}(T)\over \lambda_{2}\hbar v_{F,2}}). 
\end{equation}
Therefore, the reduced DC Josephson current for an asymmetrical junction is
\begin{equation}
{j_{1}(T)\over j_{1}(0K)}={\Delta_{1}(T)\Delta_{2}(T)\over \Delta_{1}(0)\Delta_{2}(0)} {[{\xi_{eff,1}(T)\over \lambda_{1}}+{\xi_{eff,2}(T)\over \lambda_{2}}]\over {\pi\over 2}[{\xi_{0,1}\over \lambda_{1}}+{\xi_{0,2}\over \lambda_{2}}]}. 
\end{equation}

Figure 3 shows the temperature dependence of the maximum DC supercurrent 
$j_{1}(T)$ for a Sn/Sn and a Pb/Sn junction compared with our theoretical
calculation from Eqs. (65) and (69). Data are from Fiske.$^{44}$
As in Figure 2, we used $\ell=800\AA$ and $\xi_{0}=820\AA$ for Pb and 
$\xi_{0}=1800\AA$ for Sn, respectively.
The approximate temperature dependence of $\xi_{eff}(T)$ is given by
$\xi_{eff}(T)={\pi\Delta_{0}\xi_{0}\over 2\sqrt{\pi^{2}T^{2}+\Delta^{2}}}$,
which is almost identical to the accurate numerical calculation 
of $\xi_{eff}(T)$.$^{43}$
As can be seen, the agreement between theory and experiment is fairly good.

\section{\bf Discussion } 

It is clear that more study is needed to understand the intriguing properties
of the Josephson effect. 
In particular, the sign of pair-quasi particle interference term, generalization of 
this approach to weak links and SNS junctions, and explanation of many 
unsolved problems in AC Josephson effect may be interesting problems.
For AC Josephson effect an investigation based on 
this approach will be published separately.$^{45}$

The Josephson effect in $MgB_{2}$ requires more careful study. It is highly 
desirable to fabricate clear-cut SIS junction with very low tunneling 
resistance ($R_{n}\sim 0.01-0.1\Omega$) for the detection of the Josephson 
current for the big gap. The insulating layer should have thickness not larger 
than $10\AA$ and a small band gap.

Very recently, a similar problem has been found in the flux quantization of 
superconducting cylinders.$^{46}$ Whereas most previous approaches focused 
on the phase of the effective Cooper pair wavefunction in the presence of the 
magnetic flux, flux quantization turned out to be due to the flux dependence 
of the pairing energy.$^{46}$

\section{\bf Conclusion } 
We have introduced Cooper pair wavefunction approach to the Josephson
effect in SIS tunnel junctions. We have found that the Josephson tunneling depends 
on the size of the Cooper pair and its shrinking during the tunneling. 
Accordingly there is a material-dependent threshold of tunneling resistance 
above which the DC Josephson current decreases much faster with increasing 
the tunneling resistance.
High $T_{c}$ superconductors, including $MgB_{2}$, have tightly-bound Cooper
pairs with small size, which can not tunnel through the insulating barrier 
easily, leading to extremely small critical current, in agreement with experimental findings.
This understanding also explains why the observed DC supercurrent of low 
$T_{c}$ superconductors, such as Sn and Pb, decreases much faster than 
$1/R_{n}$ above the tunneling resistances $R_{n}$ in excess of several ohms.
It is of interest that ordinary impurities limit the supercurrent by reducing
the Cooper pair size, too.
We have also shown that the (weak) temperature dependence of the Cooper pair 
size contribute to the temperature dependence of DC supercurrent.

\vspace{2pc}

\centerline{\bf ACKNOWLEDGMENTS}

Special thanks are due to Faculty of Arts and Sciences at UPR-Mayaguez for
the release time. I am grateful to Emin Yeltepe, Miguel A. Morales, 
Prof. Ju H. Kim and Prof. C. Bulutay for extensive discussions. I also 
thank Attila Altay and Profs. B. Tanatar, C. Pabon, P. Rapp, and R. Ramos for 
discussions. I take this opportunity to thank the members of Physics 
Department of Bilkent University for the hospitality extended to me during my
stay.

\vfill\eject

\begin{figure}
\caption{ Cooper pair wavefunctions near the insulating barrier. The solid lines denote the Cooper pair wavefunctions without the barrier, while the thick lines show the change of the Cooper pair wavefunctions due to the barrier.}
\end{figure}

\begin{figure}
\caption{ The maximum DC supercurrent vs the tunneling resistance $R_{n}$ for Pb-$\rm{PbO_{x}}$-Pb junction (at 4.2K) and for Sn-SnO-Pb junction (at 1.4K). Data are from Schwidtal and Finnegan, Ref. 16 and Danchi et. al., Ref. 19. Notice the crossing of the supercurrents near $R_{n}\sim 40\Omega$. The solid lines are our theoretical calculation from Eqs. (48) and (49).}
\end{figure}

\begin{figure}
\caption{ Temperature dependence of the DC supercurrent of a Sn/Sn and a Pb/Sn junction in comparison with our theoretical calculation. Data are from Fiske, Ref. 44.}
\end{figure}

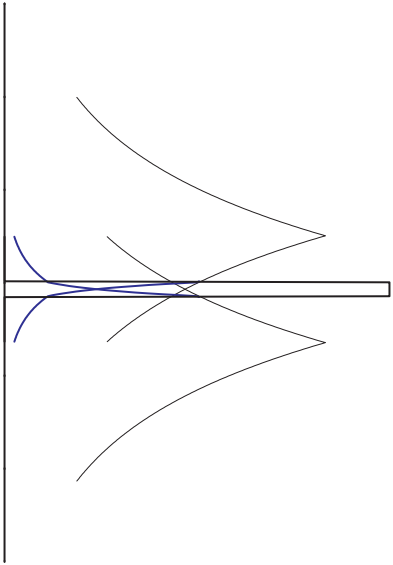
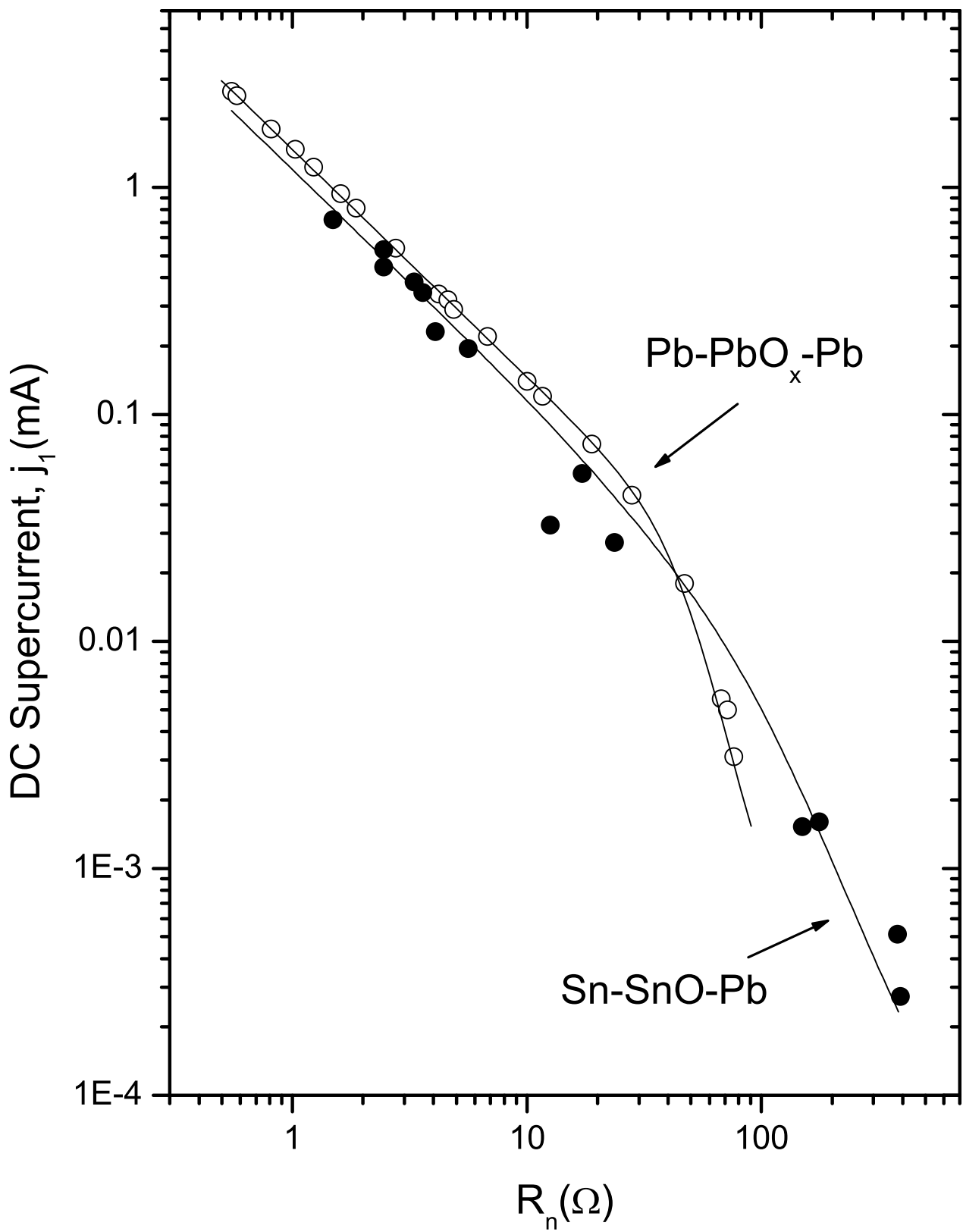
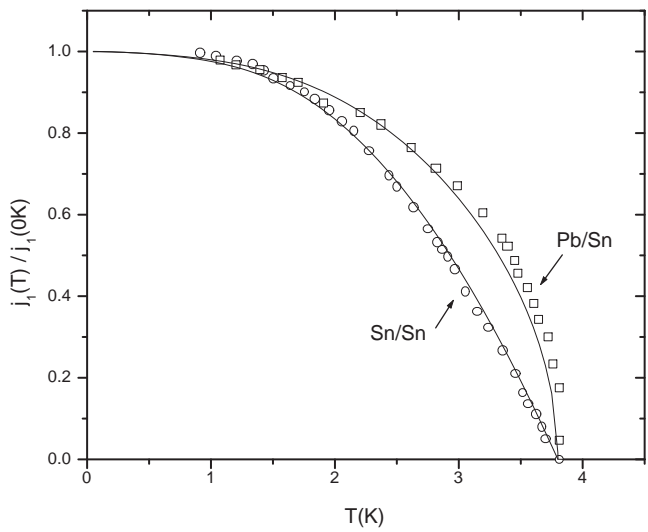
\end{document}